\documentclass[11pt]{article}
\usepackage{epsf}
\begin{document}
\title{Phase transition into superconducting mixed state  and\\ de Haas - van Alphen effect}
\author{V. P. Mineev$^{1,2}$\\$^1$\itshape Comissariat a l'Energie Atomique, \\\itshape Departament de Recherche Fondamentale sur la Matiere
Condensee, SPSMS, 38054 Grenoble, France \\\itshape 
$^2$L.D.Landau Institute for Theoretical Physics, 
Russian Academy  of Sciences, \\\itshape 117940, Moscow, Russia}
\maketitle
\begin{abstract}
 The Landau expansion for the free energy of the superconducting  
 mixed state near the 
 upper critical field in powers of the square modulus of the order 
 parameter averaged over Abrikosov lattice is derived.
 The analytical 
 calculations has been carried out in frame of Gor'kov formalism
 for 3-dimensional isotropic BCS
 model beyond the limits of quasiclassical approximation, another 
 words with Landau quantization of the quasiparticle energy levels 
 taken into consideration. The derivation is performed at
 low temperature and high enough but finite crystal's purity.
 The effect of Pauli 
 paramagnetic terms is taken into account.   The quantum 
 oscillations of the critical temperature, the order parameter's 
 amplitude and the magnetization (de Haas-van Alphen effect) 
in the mixed state are found. The 
 limitation of validity of a mean field approach due to critical 
 fluctuations (Ginzburg criterion) for the phase transition under 
 consideration is established.  
 \end{abstract}

\renewcommand {\thesection}{\arabic{section}}
\section{Introduction}
The growing body of measurements (see the most recent 
publications \cite{1,2,3,4}) demonstrates  the de Haas - van Alphen 
effect in  the mixed state of type-II superconductors. 
The theoretical description of this phenomenon is rather cumbersome 
and as a result we have a vast amount of publications 
devoted to the subject \cite{5,6,7,8,9,10,11,12,13,14,15}. 
The origination of  theoretical troubles is
a nondiagonality of  the order parameter matrix in Landau representation 
preventing a correct derivation of the 
quasiparticle energy spectrum in the mixed state.
The later can be calculated properly just near $H_{c2}$  \cite{13}.
 The intention to avoid of this problem leads to idea to develope a 
theory far below $H_{c2}$ where a vortex core radius is 
much smaller than the distance between vortices and one can assume 
the space constancy of the order parameter modulus. 
This region on the first sight intersects with field interval of
the observations of the dHvA oscillations in the mixed state extending 
down to the fields of the order of $1/2$ or even $1/5$ of 
the upper critical field (see \cite{3}). 
However in the superconductors with very large 
Ginzburg-Landau parameter $\kappa$, 
which is of the order of 20-30 in 
typical for observation of dHvA in the mixed state materials, 
the field interval of observation should still be considered as 
relating to vicinity of the upper critical field where the distance 
between the vortices  almost coincides with
the superconducting coherence length or vortex core radius. 
The former starts to exceed the later only at the fields of 
the order of $H_{c2}/\kappa$. So, in fact we need in a 
theoretical description of dHvA effect in a vicinity of the 
upper critical field  $(H_{c2}-H)/H_{c2}<1$ where one can 
assume the field inside of superconductor  as coinciding 
with homogeneous external field and  use the Abrikosov ansatz 
for the periodic in space order parameter distribution. 

Here we present the derivation of the Landau expansion 
of the free energy density in vicinity of the upper critical field  
in powers of the square modulus of the order parameter averaged 
over Abrikosov lattice, 
\begin{equation}
F_s-F_n=\alpha\Delta^2+\frac{\beta}{2}\Delta^4,
\label{e1}
\end{equation}
which is valid  until $(H_{c2}-H)/H_{c2}<1$ in spite of relative value 
of the order parameter amplitude and the distance between Landau levels
(for more exact condition of validity (\ref{e1}) see (\ref{e57})). 
The treatment of the problem is carried out in frame of the Gor'kov 
formalism with Landau quantization taken into consideration. 
It has been done earlier for  2-dimensional electron gas either numerically 
\cite{9,10,11}
or analytically   but out  of dHvA observation region\cite{12} when the
temperature $T$ exceeds the distance between the Landau levels 
\footnote{ Planck's constant $\hbar$, electron's charge $|e|$ 
and velocity of light $c$ are chosen 
to be unity throughout the paper.}
$\omega_c=H/m^*$.  The reason for the treatment of the problem in the 
unsuitable region is divergencies of $\alpha$ and $\beta$  coefficients 
in the formula (\ref{e1}) at temperature $T\to 0$ in absolutely clean limit. 
Unlike to these papers we shall treat the problem analytically for 
3-dimensional isotropic BCS model and escape of the divergencies by the 
introducing a width of Landau level $\Gamma$  originating of impurity 
scattering or a specimen's inhomogeneity.   
In addition to \cite{13} where $\alpha$ and $\beta$  coefficients
has been determined
neglecting the oscillating terms here we take them into account. 
The expressions for the functions describing the oscillating 
behaviour  of the critical temperature,  the order parameter's amplitude 
and the magnetization in the mixed state are found.
As in a normal metal the tiny oscillating corrections to the free energy 
are transformed into the big summands in the magnetic moment as the result 
of differentiation of very fast oscillating functions of the ratio of 
Fermi energy to the cyclotron frequency.
It is shown that  at the distance of the order of
$\varepsilon_F/T_c$  oscillations from 
the upper critical field  there is noticable decrease 
(of the order of unity) of the amplitude of dHvA oscillations 
of the magnetization in the mixed state. 

As was pointed out in the paper \cite{14} the Landau level splitting due to Pauli 
paramagnetic interaction of electron spin with external magnetic field  
can cause the change of character of the phase transition to  the  mixed
state.
Namely, in clean enough type-II superconductor 
the second order transition transforms into the first order transition 
at temperatures below some tricritical point depending of impurity 
concentration and determined by equation $\beta(H,T)=0$. Here we confirm
that there is a principal opportunity of such a kind phase diagramme  at
low enough temperatures $T<\Gamma$ when the Zeeman splitting  of  Landau
levels exceeds a width of them: $\mu_e H>\Gamma$. However in view of  very
complicated analytical structure of the expression for $\beta(H,T)$
it is difficult to determine definetely its sign in the region mentioned
above. One may assert only that at $\mu_e H<\Gamma$ the second order  type
transition takes place up to zero temperature.

The investigation of
the fluctuation effects is also undertaken in the present work.
Unlike to the zero field case where the fluctuation region in vicinity of 
the critical temperature is of the order of $(T_c/\varepsilon_F)^4$ that is 
negligibly small, the critical region at low temperatures in vicinity 
of the upper 
critical field  is of the order of $(T_c/\varepsilon_F)^2$. 
The period of critical temperature oscillations has the  same  order  of
magnitude. This means 
that the field interval for the reentrant normal metal-superconductor 
phase transitions originating of critical  temperature  oscillations  is
located in the critical region. Hence the  reentrant  behavior  has  the
poor chances for observation. It  is  shown  however  that  the  quantum
oscillations of $T_c(H)$ can be accessible  for  observation  in  lowest
temperature region $T<\Gamma\ll\omega_c$.

The paper is organized as follows. The formula (\ref{e1}) is derived in 
the next Section. The oscillating with magnetic field behavior of 
the $\alpha$ (or another words the oscillating behavior of the critical 
temperature) and $\beta$  coefficients are described in the following 
two Sections.  The magnetizations oscillations 
in the superconducting mixed state are calculated in Section 5. 
The critical fluctuations are discussed in the Section 6.

\section{Free energy}
The  difference of the superconductor - normal metal free  energy densities  
expanded in powers of the order parameter $\Delta({\bf r})$ and averaged 
over a superconductor volume $V$ has the form
\begin{eqnarray}
F_s-F_n=\frac{1}{V}
\int d{\bf r}(F_s({\bf r})-F_n )=
\frac{1}{gV}\int d{\bf r}|\Delta({\bf r})|^2-
\frac{1}{V}\int d{\bf r}_1d{\bf r}_2K_2({\bf r}_1,{\bf r}_2)
\Delta^*({\bf r}_1))\Delta({\bf r}_2)+ && \nonumber \\
\frac{1}{2V}\int d{\bf r}_1d{\bf r}_2{\bf r}_3d{\bf r}_4
K_4({\bf r}_1,{\bf r}_2,{\bf r}_3,{\bf r}_4)\Delta^*({\bf r}_1)
\Delta({\bf r}_2)\Delta^*({\bf r}_3)\Delta({\bf r}_4).&&
\label{e2}
\end{eqnarray}
Here $g$ is the constant of the pairing interaction.
Neglecting  all the vortex corrections that is valid
in ultra clean limit $\Gamma<\omega_c$ considered here we shall use 
for the functions $K_2$ and $K_4$ the following expressions \begin{equation}
K_2({\bf r}_1,{\bf r}_2) =\frac{1}{2}\sum_{\sigma=\pm  1}T\sum_
{\nu}G^\sigma({\bf r}_1,{\bf r}_2,\tilde \omega_\nu)
G^{-\sigma}({\bf r}_1,{\bf r}_2,-\tilde \omega_\nu),
\label{e3}
\end{equation}
\begin{equation}
K_4({\bf r}_1,{\bf r}_2.{\bf r}_3,{\bf r}_4)=
\frac{1}{2}\sum_{\sigma=\pm  1}T\sum_{\nu}G^\sigma({\bf r}_1,
{\bf r}_2,\tilde \omega_\nu)G^{-\sigma}({\bf r}_3,{\bf r}_2,-\tilde \omega_\nu)
G^\sigma({\bf r}_3,{\bf r}_4,\tilde \omega_\nu)
G^{-\sigma}({\bf r}_1,{\bf r}_4,-\tilde \omega_\nu),
\label{e4}
\end{equation}
$$
\tilde   \omega_\nu=\omega_\nu+\Gamma   sign~\omega_\nu,~~   
\omega_\nu=\pi T(2\nu+1),
$$
written by means of the electron Green functions 
$G^\sigma({\bf r}_1,{\bf r}_2,\tilde \omega_\nu)$ in normal state under an external  magnetic field. The later are related with the Green functions 
depending only on the relative coordinate 
$\tilde G^\sigma({\bf r_1}-{\bf  r_2},\tilde \omega_\nu)$ as
\begin{equation}
G^\sigma({\bf r}_1,{\bf r}_2,\tilde \omega_\nu)=
\exp \left (i\int\limits_{{\bf r}_1}^{{\bf r}_2}
{\bf A}({\bf s}) d{\bf s} \right ) 
\tilde G^\sigma({\bf r}_1-{\bf  r}_2,\tilde \omega_\nu).
\label{e5}
\end{equation}
Making use this formula one can rewright the quadratic in 
$\Delta({\bf r})$ terms in the formula (\ref{e2}) as 
\begin{equation}
\left (\frac{1}{g}-
\int d{\bf R}\exp \left (-\frac{1}{2}H \varrho^2 \right )
K_2({\bf R})\right)\frac{1}{V}\int d{\bf r} |\Delta({\bf r})|^2,
\label{e6}
\end{equation}
where $\varrho ^2 = R^2-Z^2 $ and
\begin{equation}
K_2({\bf R})=\frac{1}{2}\sum_{\sigma=\pm 1}T\sum_{\nu }
\tilde G^\sigma({\bf R},\tilde \omega_\nu)\tilde G^{-\sigma}({\bf R},-\tilde \omega_\nu)).
\label{e7}
\end{equation}

As it was discussed in Introduction 
for the strong type-II superconducting materials  
the Abrikosov solution for the order parameter 
\begin{equation}
\Delta({\bf r})=\Delta f({\bf r}),
\label{e8}
\end{equation}
\begin{equation}
f({\bf r})=\sqrt[4]{2}\sum_{\nu=integer}\exp\left(i\frac{2\pi\nu}{a}y-
\left(\frac{x}{\lambda}+\frac{\pi\nu}{a}\lambda\right)^2\right).
\label{e9}
\end{equation}
is valid in broad enough vicinity of the upper critical field. 
Here for simplicity we have chosen the Abrikosov square lattice 
solution in the Landau gauge ${\bf A}({\bf r})=(0,Hx,0)$. 
The elementary cell edge length $a$ is such that $a^2=\pi\lambda^2$, 
where $\lambda=H^{-1/2}$ is magnetic length. 
We assume that magnetic field is uniform and coincides with 
an external field. This is certainly true in vicinity of $H_{c2}$  
for  the superconductors with large Ginzburg-Landau parameter . 

Using formulas (\ref{e6}),(\ref{e8}), (\ref{e9}) one can rewright (\ref{e2}) 
in the shape (\ref{e1})
$$
F_s-F_n=\alpha\Delta^2+\frac{\beta}{2}\Delta^4,
$$
where
\begin{equation}
\alpha=\frac{1}{g}-
\int d{\bf R}\exp \left (-\frac{1}{2}H \varrho^2 \right )K_2({\bf R})
\label{e10}
\end{equation}
and
\begin{equation}
\beta=\frac{1}{V}\int d{\bf r}_1d{\bf r}_2{\bf r}_3d{\bf r}_4
K_4({\bf r}_1,{\bf r}_2,{\bf r}_3,{\bf r}_4)f^*({\bf r}_1)
f({\bf r}_2)f^*({\bf r}_3)f({\bf r}_4).
\label{e11}
\end{equation}
The amplitude of the order parameter $\Delta$ has to be found from the 
free energy minimum condition.  

The $\alpha$ and $\beta$ coefficients are the oscillating functions 
of the magnetic field which will be obtained in the following two Sections.

\section{Critical temperature oscillations }
The critical temperature as a function of magnetic field at $T\ll T_c$
have been found in the paper by L.Gruenberg and L.Guenther \cite{16}. 
It is determined by the equation 
\begin{equation}
\alpha(H,T)=0,
\label{e12}
\end{equation}
where
\begin{equation}
\alpha(H,T)=\frac{N_o}{2}\left\{\frac{H-H_{c2o}}{H_{c2o}}+
\frac{T^2}{T_c^2}S_0-
2\pi^{3/2}\left(\frac{\omega_c}{\mu}\right)^{1/2}S_1-
2^{3/2}\pi^{1/2}\frac{\omega_c}{\mu}S_2\right\},
\label{e13}
\end{equation}
\begin{equation}
S_1=\frac{2\pi T}{\omega_c}\Re\sum_{n=1}^{\infty}\sum_{\nu=0}^{\infty}
\exp\left(-\frac{4\pi n(\tilde \omega_\nu+i\mu_e H)}{\omega_c}\right),
\label{e14}
\end{equation}
\begin{equation}
S_2=\frac{2\pi T}{\omega_c}\Re\sum_{n=1}^{\infty}\sum_{m>n}^{\infty}
\sum_{\nu=0}^{\infty}
(-1)^{m+n}
\exp\left(-\frac{2\pi(m+n)(\tilde \omega_\nu+i\mu_e H)}{\omega_c}\right)
\frac{\cos
\left(2\pi|n-m|\frac{\mu}{\omega_c}-\frac{\pi}{4}\right)}
{|n-m|^{1/2}},
\label{e15}
\end{equation}
\begin{equation}
H_{c2o}=\frac{\pi^2 {\it e}^2}{2\gamma}\frac{T_c^2}{v_F^2}.
\label{e16}
\end{equation}
Here, $\mu$ is chemical potential, $v_F$ is Fermi velocity,
$\Gamma$ is Landau level width, 
$\mu_e$ is the electron's magnetic  moment,  $N_0=m*k_F
/2\pi^2$  is  the  normal  metal
electron density of states per  one  spin  projection.  $S_0$  is  slow
(logariphm) function of magnetic field with numerical value of the order
of unity.

The substitution of the
first two terms of (\ref{e13}) in the equation (\ref{e13}) 
reproduces of the 
quasiclassical result found by L.Gor'kov \cite{17}
\begin{equation}
H=H_{c2o}(1-\frac{T^2}{T_c^2}S_0) 
\label{e17}
\end{equation}
for the upper critical field 
at temperature tending to zero . The third and the fourth terms 
in the formula (\ref{e13}) give 
correspondingly    nonoscillating    and    oscillating      corrections
to Gor'kov's expression owing
to quantization of the quasiparticle levels. 
For the further considerations it will be convinient  to  represent  the
functions $S_1$ and $S_2$ in more simple analytical form.

The summation over $\nu$ in (\ref{e14}) yields 
\begin{equation}
S_1=\frac{\pi T}{\omega_c}\Re\sum_{n=1}^{\infty}
\frac{\exp\left(-\frac{4\pi n a}{\omega_c}\right)}
{\sinh \frac{4\pi^2nT}{\omega_c }} ,
\label{e18}
\end{equation}
where
\begin{equation}
a=\Gamma+i\eta\omega_c,
\label{e19}
\end{equation}
and
\begin{equation}
\eta\omega_c=\mu_e H-\frac{q}{2}\omega_c,
\label{e20}
\end{equation}
is the Zeeman energy  deviation from a  half  integer  number  $q/2$  of
distances $\omega_c$ between Landau levels. For noninteracting  electron
gas in a perfect crystal $a=0$ and
\begin{equation}
S_1\approx \frac{1}{4\pi}\ln \frac{\omega_c}{4\pi^2T} 
\label{e21}
\end{equation}
diverges at $T\to  0$.  This  result  formally  means  an  existance  of
superconductivity of noninteracting electron gas  in  a perfect  crystal  at
$T=0$ in arbitrary large field \cite{16}.

In the opposite limit
$\pi T\ll |a|$ when
\begin{equation}
S_1=\Re\sum_{n=1}^{\infty}
\left (\frac{1}{4\pi n}-\frac{2\pi^3n}{3}\left(
\frac{T}{\omega_c}\right)^2\right)
\exp\left(-\frac{4\pi na}{\omega_c}\right).
\label{e22}
\end{equation}
At $|a|\ll \omega_c/4\pi$ it is
\begin{equation}
S_1=\frac{1}{4\pi}\ln \frac{\omega_c}{4\pi|a|}-\frac{\pi}{24}\left (
\frac{T}{|a|}\right )^2.
\label{e23}
\end{equation}

For $4\pi\Gamma\sim \omega_c$ one can estimate sum (\ref{e18})
by its first term
\begin{equation}
S_1=\frac{\pi T}{\omega_c\sinh \frac{4\pi^2T}{\omega_c }}
\exp\left(-\frac{4\pi \Gamma}{\omega_c}\right)\cos4\pi\eta.
\label{e24}
\end{equation}
 The substitution (\ref{e24}) into equations (\ref{e13}), (\ref{e12}) 
 provides us by the averaged over oscillations 
behavior  of  upper
critical field at low temperatures 
\begin{equation}
\bar {H}_{c2}(T)=H_{c2o}\left \{1+\frac{1}{2}\sqrt{ \frac{\pi\omega_{c2}}{\mu}}
\left(1-\frac{8\pi^4}{3}\left(
\frac{T}{\omega_{c2}}\right )^2\right)\exp\left(-\frac
{4\pi\Gamma}{\omega_{c2}}\right)\cos4\pi\eta\right\}
\label{e25}
\end{equation}
Here $\omega_{c2}=H_{c2}/m^*$ is the  cyclotron  frequency  
at  the  upper  critical field. Let us  denote the
number of Landau levels   at $H=H_{c2}$ disposed below the Fermi level 
$\varepsilon_F$ as 
\begin{equation}
n_{c2}=
\frac{\varepsilon_F}{\omega_c}=\frac{1}{2\pi^2}\left(\frac{\varepsilon_F}
{T_c}\right)^2.
\label{e26}
\end{equation}
 The number $n_{c2}$
in typical materials with small Fermi energy and relatively high $T_c$
where dHvA effect in the mixed state has been observed is of  the  order
of one hundred.
So, the upper critical field at zero temperature is noticably larger
$(\sim n_{c2}^{-1/2})$ than its quasiclassical value (\ref{e16}).
The decreasing $H_{c2}$ with temperature follows $T^2$ law. 
This parabolic dependence in the region 
$$
T\ll \frac{\omega_c}{2\pi^2}\sim \frac {T_c^2}{\varepsilon_F}
$$
is much faster than Gorkov's parabola  (\ref{e17}).

For $S_2$ the summation over $\nu$ in (\ref{e14}) produces
\begin{equation}
S_2=\frac{\pi T}{\omega_c}\Re\sum_{n=1}^{\infty}\sum_{m>n}^{\infty}
(-1)^{m+n}
\frac{\exp\left(-\frac{2\pi(m+n)(\Gamma+i\mu_e H)}{\omega_c}\right)}
{\sinh \frac{2\pi^2(m+n)T}{\omega_c }}\frac{\cos \left(2\pi(m-n)
\frac{\mu}{\omega_c}-\frac{\pi}{4}\right)}{(m-n)^{1/2}}.
\label{e27}
\end{equation}
Changing the summation variables to $n$ and $m-n=l$ we get
\begin{equation}
S_2=\frac{\pi T}{\omega_c}\Re\sum_{l=1}^{\infty}
\frac{(-1)^l}{\sqrt l}
\exp\left(-\frac{2\pi l(\Gamma+i\mu_e H)}{\omega_c}\right)
\cos\left(2\pi l\frac{\mu}{\omega_c}-\frac{\pi}{4}\right)
\sum_{n=1}^{\infty}
\frac{\exp\left(-\frac{4\pi n(\Gamma+i\mu_e H)}{\omega_c}\right)}
{\sinh \frac{2\pi^2(l+2n)T}{\omega_c }}.
\label{e28}
\end{equation}
Similar to $S_1$ for noninteracting electron gas in a perfect crystal 
($(2\mu_e H=\omega_c,~ \Gamma=0)$) the expression for $S_2$ is
diverged at $T\to 0$  (see also \cite{16}). 

In more realistic case $4\pi\Gamma\sim\omega_c$
one can keep only the first term in the sum over $n$:
\begin{equation}
S_2=\frac{\pi T}{\omega_c}\sum_{l=1}^{\infty}
\frac{(-1)^l\exp\left(-\frac{2\pi\Gamma(l+2)}{\omega_c}\right)}
{\sqrt l~\sinh\frac{2\pi^2(l+2)T}{\omega_c}}
\cos\left(2\pi l\frac{\mu}{\omega_c}-\frac{\pi}{4}\right)
\cos\left(\frac{2\pi\mu_e H(l+2)}{\omega_c}\right).
\label{e29}
\end{equation}

Substitution of (\ref{e24}) and (\ref{e29})  into  (\ref{e12})  and  the
resolution of the equation $\alpha(H,T)=0$ in respect of $T$ gives the
oscillating behavior of critical temperature. The simple estimation
shows     that      the      amplitude      of      oscillations      at
$T\sim\omega_{c2}\sim2\pi^2T_c^2/\varepsilon_F$ is of the order of
\begin{equation}
\frac{\delta T_{osc}}{T_c}\sim\left(\frac{T_c}{\varepsilon_F}\right)^2
\label{e30}
\end{equation}
At lower temperatures the amplitude of oscillations is somethat larger.
The period of one oscillation $\delta H$ is easy to find from
$$
1=\delta n=\delta\frac{\varepsilon_F}{\omega_c}=
-\frac{\varepsilon_F}{\omega_c}\frac{\delta H}{H}
$$
So,
\begin{equation}
\frac{\delta H}{H_{c2}}\approx2\pi^2\left(\frac{T_c}{\varepsilon_F}\right)^2.
\label{e31}
\end{equation}
As we shall see later this value coincides with the critical region
in vicinity of upper critical  field  at  low  temperatures.  Hence  the
critical temperature oscillating behavior as a function of magnetic
field exists only in frames of mean field approximation  and  has  poor
chances to be observable. On the other hand outside the narrow region of
strong critical fluctuations in  the  mixed  superconducting  state  the
function
\begin{equation} 
\alpha(H,T)=\bar {\alpha}(H,T)+\alpha^{osc}(H,T)
\label{e32}
\end{equation}
consists of smooth function
\begin{equation} 
\bar {\alpha}(H,T)=\frac{N_0}{2}\frac{H-\bar {H}_{c2}(T)}{H_{c2o}}
\label{e33}
\end{equation}
(see (\ref{e25})) and
fast oscillating function of magnetic field 
\begin{equation} 
\alpha^{osc}(H,T)=-2^{1/2}\pi^{3/2}N_0\frac{T}{\mu}\sum_{l=1}^{\infty}
\frac{(-1)^l\exp\left(-\frac{2\pi\Gamma(l+2)}{\omega_c}\right)}
{\sqrt l~\sinh\frac{2\pi^2(l+2)T}{\omega_c}}
\cos\left(2\pi l\frac{\mu}{\omega_c}-\frac{\pi}{4}\right)
\cos\left(\frac{2\pi\mu_e H(l+2)}{\omega_c}\right)
\label{e34}
\end{equation}
giving the main
contribution to the oscillating part of the magnetization. 

\section{ Oscillations of $\beta(H,T)$}
Instead of formula (\ref{e11}) it is more suitable
to use an expression for  $\beta$  in  the so  called  magnetic  sublattices
representation (see \cite{13}) 
\begin{equation}
\beta(H,T)=\frac{1}{2}\sum_{\sigma=\pm 1}T\sum_{ \nu}\sum_{n,n',m,m'}\int 
\frac{dk_z}{2\pi}G^{-\sigma}(\xi_{n k_z},-\tilde \omega_\nu)
G^{\sigma}(\xi_{m k_z},\tilde \omega_\nu)
G^{-\sigma}(\xi_{n' k_z},-\tilde \omega_\nu)G^{\sigma}(\xi_{m' k_z},
\tilde\omega_\nu)F_{mm'}^{nn'},
\label{e35}
\end{equation}
where
\begin{equation}
G^\sigma(\xi_{n k_z},\tilde \omega_{\nu})=
\frac{1}{i\tilde \omega_{\nu}-\xi_{n k_z}  + \sigma\mu_eH},
\label{e36}
\end{equation}
\begin{equation}
\xi_{n k_z}=\omega_c(n+\frac{1}{2})+\frac{k_z^2}{2m^*}-\mu, 
\label{e37}
\end{equation}
\begin{equation}
F_{mm'}^{nn'}=\int \frac{d\vec {\bf q}}{(2\pi)^2}
f_{nm}(\vec {\bf q})f_{nm'}^*(\vec {\bf q})f_{n'm}(\vec {\bf q})f_{n'm'}^*
(\vec {\bf q})
\label{e38}
\end{equation}
and $f_{nm}(\vec {\bf q})$ are the matrix elements of functions (\ref{e9}).

In view of extremely complicated structure of
general  formula  (\ref{e35}) we  are
limited ourselves by the analysis of diagonal terms $n=n'=m=m'$
giving the main contribution to the expression (\ref{e35}) 
at $\Gamma<\omega_c$. For this case
as was shown in the paper \cite{13}
$F_{nn}^{nn}\approx 1/(2\pi\lambda)^2 n$ and
\begin{equation}
\beta(H,T)=\frac{1}{2(2\pi\lambda)^2}\sum_{\sigma=\pm 1}
T\sum_{ \nu}\sum_{n=0}^{\infty}\int 
\frac{dk_z}{2\pi n}
\left(G^{\sigma}(\xi_{n k_z},\tilde \omega_\nu)
G^{-\sigma}(\xi_{n k_z},-\tilde \omega_\nu)\right )^2.
\label{e39}
\end{equation}
Let us represent 
\begin{equation}
\beta(H,T)=\bar {\beta}(H,T)+\beta^{osc}(H,T)
\label{e40}
\end{equation}
as the sum  of  smooth  nonoscillating  function  and  fast  oscillating
function of magnetic field.
For the calculation of nonoscillating part of $\beta$
following the papers \cite{13,14} 
one can substitute the summation over $n$ by the integration according to
$$
\frac{1}{2\pi\lambda^2}\sum_{n=0}^{\infty}\int\frac{dk_z}{2\pi}=
\int_{0}^{\infty}\frac{dn}{2\pi\lambda^2}\int\frac{dk_z}{2\pi}=
N_0\int d\xi\int_0^{\pi/2}\sin \theta d\theta
$$
Making use $n\approx \mu\sin^2\theta/\omega_c$ and performing the integration 
over $\theta$ we get
\begin{equation}
\bar {\beta}(H,T)=\frac{\omega_c\ln{\frac{\mu}{\omega_c}}}{8\pi\mu}
\sum_{\sigma=\pm 1}
T\sum_{\nu}N_0  \int  d\xi~
\left (G^\sigma(\xi,\tilde \omega)
G^{-\sigma}(\xi,-\tilde \omega)\right )^2,
\label{e41}
\end{equation}

The integration over $\xi$ yields
\begin{equation}
\bar {\beta}(H,T)=N_0\frac{\omega_c\ln {\frac{\mu}{\omega_c}}}{4 \mu}
T\sum_{\nu=0}^{\infty}
\frac{\tilde \omega^3-3\tilde \omega(\mu_e H)^2}
{(\tilde \omega^2+(\mu_e H)^2)^3},
\label{e42}
\end{equation}
This expression can be positive or negative depending on  the relation
between the values of $\Gamma$ and $\mu_e H$.  At  $T=0~K$ when it  is
possible to change the summation to the integration we obtain
\begin{equation}
\bar{\beta}(H,0)=N_0\frac{\omega_c\ln{\frac{\mu}{\omega_c}}}{16\pi\mu}
\frac{\Gamma^2-(\mu_e H)^2}{(\Gamma^2+(\mu_e H)^2)^2},
\label{e43}
\end{equation}
This result demonstrates the positiveness of $\bar\beta$ up to
zero  temperature at $ \Gamma>\mu_e  H$.
In the opposite case $\bar\beta$ at $T=0~K$ is negative. 
Hence below some temperature $\sim \mu_e H$ the second order type 
transition to the superconducting state starts to be of the first order. 
Becouse we have thrown out the nondiagonal terms it is difficult 
to say definetely is it the case or not and if 
it is so what is the temperature  of  tricritical  point.  
Leaving this problem for future experimental and theoretical investigations 
let us assume positive value of 
\begin{equation}
\bar {\beta}(H,0) \approx N_0 \frac{\omega_c\ln \frac{\mu}{\omega_c}}
{16\pi\mu\Gamma^2},
\label{e44}
\end{equation}
which is valid at least at the fulfillment  of the conditions
$T,~\mu_e H<\Gamma<\omega_c$.

To find the oscillating part of $\beta(H,T)$ at the same conditions 
let us apply  to the expression (\ref{e37}) the Poisson summation formula 
\begin{eqnarray}
&&\beta^{osc}(H,T)= \nonumber \\
&&N_0\frac{\omega_c}{2\sqrt {2m*\mu}}\Re\sum_{\sigma=\pm 1}T\sum_{\nu}
\int^{\infty}d\varepsilon \frac{\partial n}{\partial \varepsilon}
\int \frac{dk_z}{2\pi n(\varepsilon,k_z)}
\sum_{l=1}^{\infty}e^{2\pi inl}
\left(G^{\sigma}(\varepsilon-\mu,\tilde \omega_\nu)
G^{-\sigma}(\varepsilon-\mu,-\tilde \omega_\nu)\right )^2,
\label{e45}
\end{eqnarray}
here $\varepsilon=\xi_{n,k_z}+\mu$.
Subsequent calculations follows by well known Lifshits, Kosevich
derivation \cite{18} of normal metal de Haas - van Alphen oscillations.
Performing the saddle point integration over $k_z$ we get
\begin{eqnarray}
\beta^{osc}(H,T)= 
N_0\frac{\omega_c^{3/2}}{2^{5/2}\pi\mu^{3/2}}\Re\sum_{\sigma=\pm 1}
T\sum_{\nu}\sum_{l=1}^{\infty}\frac{(-1)^l}{\sqrt l}
\exp {\left  (2\pi  il\frac{\mu}{\omega_c}-i\frac{\pi}{4}\right )}\times&&\nonumber \\
\int^{\infty}_{-\infty}d\xi\frac{ \exp {\left  (2\pi il
\frac{\xi}{\omega_c}\right )}} {(i\tilde \omega_\nu-\xi+\sigma\mu_e H)^2
(-i\tilde \omega_\nu-\xi-\sigma\mu_e H)^2}.&&
\label{e46}
\end{eqnarray}
The integration over $\xi$ yields
\begin{eqnarray}
\beta^{osc}(H,T)= 
N_0&\!\!\!\frac{\omega_c^{3/2}}{(2\mu)^{3/2}}
\sum_{l=1}^{\infty}\frac{(-1)^l}{\sqrt l}
\cos {\left (2\pi l\frac{\mu}{\omega_c}-\frac{\pi}{4}\right )}&\!\!\!
\Re T\sum_{\nu=0}^{\infty}
\exp {\left  (-2\pi l\frac{\tilde \omega_\nu+i\mu_e H}
{\omega_c}\right )}\nonumber \\
\!\!\!&\times \left \{\frac{1}{(\tilde \omega_\nu+i\mu_e H)^3}+
\frac{2\pi l}{\omega_c(\tilde \omega_\nu+ i\mu_e H)^2}\right \}.&
\label{e47}
\end{eqnarray}
At the fulfilment of the conditions $T, \mu_e H <\Gamma <\omega_c$
one can rewrite this expression in more simple form:
\begin{equation}
\beta^{osc}(H,0)= N_0\frac{\omega_c^{3/2}}{2\pi (2\mu)^{3/2}\Gamma^2}
\sum_{l=1}^{\infty}\frac{(-1)^l}{\sqrt l}
\cos {\left (2\pi l\frac{\mu}{\omega_c}-\frac{\pi}{4}\right )}
\exp {\left  (-\frac{2\pi l\Gamma}{\omega_c}\right )}
I\left (\frac{2\pi l\Gamma}{\omega_c}\right ),
\label{e48}
\end{equation}
where
$$
I(x)=
\int_0^{\infty} dy\left \{\frac{1}{(y+1)^3}+
\frac{x}{(y+1)^2}\right \}
\exp {(-xy)}.
$$
Thus, as it is expected, the oscillating 
part of $\beta$ (\ref{e47}), (\ref{e48}) is $\sqrt n_{c2}$ times 
smaller than its smoth nonoscillating part (\ref{e43}), (\ref{e44}).

The preliminary calculations of $\beta$ has been published in \cite{19}.
Although the results of that paper are in general qualitatively correct
there there are the errors in the values of $\bar \beta$ and
$\beta^{osc}$ (compare formula (\ref{e6}) in \cite{18} and (\ref{e44}),
(\ref{e48}) in the present article).
\section{Magnetizations oscillations }
Minimization of the expression (\ref{e1}) over $\Delta$ yields
\begin{equation}
F_s=F_n-\frac{\alpha^2}{2\beta}.
\label{e49}
\end{equation}
Now we can calculate the magnetization
\begin{equation}
M_s=-\frac{\partial F_s}{\partial H}.
\label{e50}
\end{equation}
Keeping at the differentiation only the fast oscillating terms we get
\begin{equation}
M_s^{osc}\simeq M_n^{osc}+
\frac{\bar {\alpha}}{\bar {\beta}}
\frac{\partial \alpha^{osc}}{\partial H}-
\frac{\bar {\alpha}^2}{2\bar {\beta}^2}
\frac{\partial \beta^{osc}}{\partial H}.
\label{e51}
\end{equation}
Here $M_n^{osc}$ is the  oscillating  part  of  normal  metal  magnetization
\cite{18} including Dingle factor $\exp(-2\pi l\Gamma/\omega_c)$ due 
to scattering by imperfections
\begin{equation}
M_n^{osc}=\sum_{l=1}^{\infty} M_l.
\label{e52}
\end{equation}
Here
\begin{equation}
M_l=N_0\frac{\omega_c^2}{2^{1/2}\pi H}\left
(\frac{\mu}{\omega_c}\right)^{1/2}\frac{(-1)^{l+1}
\lambda_l}{l^{3/2}\sinh \lambda_l}
\sin \left (2\pi l\frac{\mu}{\omega_c}-\frac{\pi}{4}\right )
\cos \left (2\pi l\frac{\mu_e H}{\omega_c}\right )
\exp \left (-2\pi l\frac{\Gamma}{\omega_c}\right ),
\label{e53}
\end{equation}
\begin{equation}
\lambda_l=\frac{2\pi^2 lT}{\omega_c}.
\label{e54}
\end{equation}
For  the  subsequent  terms  in  (\ref{e51})  making  use   (\ref{e33}),
(\ref{e34}) and approximative expressions (\ref{e44}) and
(\ref{e48}) for the smooth and oscillating parts of $\beta$ we get
\begin{eqnarray}
&&\frac{\bar {\alpha}}{\bar {\beta}}
\frac{\partial \alpha^{osc}}{\partial H}=
-N_0\frac{2^{7/2}\pi^{3/2}\Gamma^2}{H}
\frac{\mu}{\omega_c\ln \frac{\mu}{\omega_c}}
\frac{\bar H_{c2}(T)-H}{H_{c2o}} \nonumber\times \\
&& \sum_{l=1}^{\infty}\frac{(-1)^{l+1}l^{1/2}\lambda_{l+2}}
{(l+2)\sinh \lambda_{l+2}}
\sin \left (2\pi l\frac{\mu}{\omega_c}-\frac{\pi}{4}\right )
\cos \left (2\pi (l+2)\frac{\mu_e H}{\omega_c}\right )
\exp \left (-2\pi (l+2)\frac{\Gamma}{\omega_c}\right ),
\label{e55}
\end{eqnarray}
\begin{eqnarray}
&&-\frac{\bar {\alpha}^2}{2\bar {\beta}^2}
\frac{\partial \beta^{osc}}{\partial H}= 
N_0\frac{2^{5/2}\pi^2\Gamma^2}{H}
\frac{\mu^{3/2}}{\omega_c^{3/2}\ln^2{ \frac{\mu}{\omega_c}}}
\left (\frac{\bar H_{c2}(T)-H}{H_{c2o}}\right )^2 \times\nonumber \\
 &&\sum_{l=1}^{\infty}(-1)^{l+1}l^{1/2}
\sin \left (2\pi l\frac{\mu}{\omega_c}-\frac{\pi}{4}\right )
\exp \left (-2\pi l\frac{\Gamma}{\omega_c}\right )
I \left (2\pi l\frac{\Gamma}{\omega_c}\right ).
\label{e56}
\end{eqnarray}
The  corrections  to  the
dHvA amplitude in the mixed superconducting state have the same structure
as the normal metal part: each term in the sums (\ref{e55}), (\ref{e56})
oscillates  with the  frequency  $l\mu/\omega_c$  corresponding  to  the
extremal crossection of the Fermi surface.   
From these expressions at $4\pi\Gamma\sim\omega_c$ it is easy to see 
that until
\begin{equation}
\frac{\bar H_{c2}(T)-H}{H_{c2o}}~<~\sqrt {\frac{\omega_c}{\mu}}
\ln {\frac{\mu}{\omega_c}}
\label{e57}
\end{equation} 
the following inequality takes place
\begin{equation}
|M_n^{osc}|~>~\left |\frac{\bar {\alpha}}{\bar {\beta}}
\frac{\partial \alpha^{osc}}{\partial H}\right |~>~
\left |-\frac{\bar {\alpha}^2}{2\bar {\beta}^2}
\frac{\partial \beta^{osc}}{\partial H}\right |.
\label{e58}
\end{equation} 
So, in the limit of validity of the condition (\ref{e57}) one can neglect
of the third term in the formula (\ref{e51}) and treat the (\ref{e55})
as the main mixed state correction to the oscillating part
of magnetization in the normal state. Summing up (\ref{e52}) and
(\ref{e55}) we obtain 
\begin{equation}
M_s^{osc}=M_n^{osc}+\frac{\bar {\alpha}}{\bar {\beta}}
\frac{\partial \alpha^{osc}}{\partial H}= 
\sum_{l=1}^{\infty}M_lM_{sl},
\label{e59}
\end{equation} 
\begin{equation}
M_{sl}=1-\frac{1}{\sqrt\pi}\left (\frac{4\pi\Gamma}{\omega_c}\right )^2
\frac{\mu^{1/2}}{\omega_c^{1/2}\ln \frac{\mu}{\omega_c}}
\frac{\bar H_{c2}(T)-H}{H_{c2o}}
\frac{l^2\lambda_{l+2}\sinh \lambda_l}
{(l+2)\lambda_l\sinh \lambda_{l+2}}
\frac{\cos \left (2\pi (l+2)\frac{\mu_e H}{\omega_c}\right )}
{\cos \left (2\pi l\frac{\mu_e H}{\omega_c}\right )}
\exp\left (-\frac{4\pi\Gamma}{\omega_c}\right ),
\label{e60}
\end{equation}
This expression presents the  main  result  of  the  paper.  Looking  on
(\ref{e60}) one can conclude that at $\mu_e H\approx\omega_c/2$ as  well
as at  $\mu_e H\ll\omega_c/2$ the amplitude of  de  Haas  -  van  Alphen
effect in the superconducting mixed state is less  than  in  the  normal
state. For the intermediate values of Zeeman splitting like
$\mu_e  H\approx\omega_c/4$   the   amplitude   of   the   magnetization
oscillations in the mixed state can be even larger than  in  the  normal
state.

The region of validity  of  the  result (\ref{e59}) is  determined  by  the
inequality (\ref{e57})  
which can be rewritten as (see (\ref{e26}))
\begin{equation}
\frac{\bar H_{c2}(T)-H}{H_{c2o}}~<~\frac{\ln n_{c2}}{\sqrt {n_{c2}}}. 
\label{e61}
\end{equation} 
If we remember that period of oscillations (\ref{e31}) is of the order of  
\begin{equation}
\frac{\delta H}{H_{c2o}}\approx\frac{1}{n_{c2}} 
\label{e62}
\end{equation} 
that means the formula (\ref{e59}) describes the  oscillating  part
of magnetization in the field interval corresponding to 
\begin{equation}
\sqrt {n_{c2}}\ln n_{c2}. 
\label{e63}
\end{equation}
oscillations below $H_{c2}$. This number in the typical
for observation dHvA effect in the superconducting mixed
state materials is of the order of several tens. Out this field interval
the corrections to the normal metal amplitude  of any order
$\sim (H_{c2}-H)^n$ shall be the same order of magnitude and the result   
(\ref{e59}) is inapplicapable to the mixed state dHvA effect description.
It is worth to be noted here that the region (\ref{e63}) coincides  with
the region of the existance of gapless superconductivity found in the paper
\cite{13}. So, out the field interval (\ref{e57}) below upper critical field
the de Haas van Alphen effect in the superconducting mixed state has to
be negligibly small.
\section {Critical fluctuations}
The problem of critical behavior of type II superconductors  near  the upper
critical field had been considered in many theoretical papers (see for 
instance \cite{20} and references therein). We shall not be interested here
by the solution of this problem as whole but just 
an estimation of a width of the fluctuational region at very low temperatures
in very clean materials.  To  establish  the  region  of  importance  of
critical fluctuations under magnetic field near the zero temperature
let us remind first how to estimate it   in  zero  magnetic  field  near
$T_c$. The most convinient for  our  purposes  way  is  to  compare  the
density   of   energy   of    critical    fluctuations    $F_{fl}\sim
T/(\xi(T))^3$    with  mean  field  energy
density $F_{mf}\sim N_0(T-T_c)^2$ above  critical temperature.
Here 
$$
\xi(T)=\xi_0\sqrt{\frac{T_c}{T-T_c}}
$$  
is   the   coherence   length.   The
comparison of these values leads us to well  known  condition  (Ginzburg
criterion)
\begin{equation}
\frac{T-T_c}{T_c}~<~\left (\frac{T_c}{\varepsilon_F}\right )^4 
\label{e64}
\end{equation}
of importance of critical fluctuations. To  get  the  similar  condition
near $H_{c2}$  at low temperatures one needs to take into account the
one dimentional character of fluctuations along magnetic field direction
such that the density of energy of critical fluctuations will be 
\begin{equation}
 F_{fl}\sim \frac{T}{2\pi \lambda^2\xi(H)}.
\label{e65}
\end{equation}
Where $\lambda=H^{-1/2}$ is magnetic length coinciding at $H=H_{c2}(T=0)$
with zero temperature coherence length \\
$\xi_0=v_F/2\pi  T_c$.  The  field  dependent  coherence  length   above
$H_{c2}$ is given by  
\begin{equation}
\xi(H)\approx\lambda\sqrt{\frac{H_{c2}}{H-H_{c2}}}.
\label{e66}
\end{equation}  
Here we have omitted the  logariphmic
corrections to $\xi(H)$ (see \cite{21}).

The density of mean field free energy has the form 
\begin{equation}
F_{mf}=\frac{\bar {\alpha}^2}{2\bar {\beta}}=
2\pi N_0\frac{\varepsilon_F\Gamma^2}
{\omega_c\ln \frac{\varepsilon_F}{\omega_c}}
\left (\frac{H-H_{c2}}{H_{c2}}\right )^2
\label{e67}
\end{equation}  

The comparison of equations (\ref{e65}) and (\ref{e67}) shows
that the  critical  fluctuations  contribution  to  the  thermodynamical
values surpass the corresponding mean field values at  
\begin{equation}
\frac{H-H_{c2}}{H_{c2}}~<~\pi^2
\left (\frac{T_c}{\varepsilon_F}\right )^2
\left (\frac{T\omega_c}{\Gamma^2}
\ln {\frac{\varepsilon_F}{\omega_c}}\right )^{2/3}
\label{e68}
\end{equation}  
This  inequality  establishes  the  region  of  importance  of  critical
fluctuations (Ginzburg criterion) above $H_{c2}$ at low temperatures.
The applications of this criterion has its own limitations. The point is
that it is valid for classical or thermal fluctuations 
with energies $\sim T$ larger than
minimal frequency of the order parameter fluctuations
\begin{equation}
\frac{H-H_{c2}}{H_{c2}}~<~\frac{T}{T_c}.
\label{e69}
\end{equation}
When temperature decreases, the region of importance of thermal flugtuations
(\ref{e68}) shrinks  slower  ($\propto  T^{2/3}$)  
than  the  region (\ref{e69})
of applicability of Ginzburg criterion. So, at lowest  temperatures  the
formula (\ref{e68}) is inapplicable. Let us compare vicinities (\ref{e68})
and  (\ref{e69})  say  at  
\begin{equation}
T\sim  \Gamma  \sim  \frac{\omega_{c2}}{4\pi}
\sim\frac{\pi T_c^2}{2\varepsilon_F}. 
\label{e70}
\end{equation}
For the region
of importance of critical fluctuations (\ref{e68}) we get 
\begin{equation}
\frac{H-H_{c2}}{H_{c2}}~<~\left 
(4\pi^4\ln {\frac{\varepsilon_F}{\omega_c}}\right )^{2/3}
\left (\frac{T_c}{\varepsilon_F}\right )^2
\label{e71}
\end{equation}
and the thermal fluctuations region (\ref{e69}) is limited by
\begin{equation}
\frac{H-H_{c2}}{H_{c2}}~<~ \frac{\pi}{2}
\frac{T_c}{\varepsilon_F}.
\label{e72}
\end{equation}
We see that the region (\ref{e71}) is still parametrically more narrow than 
the region (\ref{e72}). However due to the large numerical factor in
(\ref{e71}) these regions can be proved of the same order of magnitude.

At lower temperatures we have to calculate the quantum fluctuation
contribution. For the problem under discussion this contribution is 
exponentially small, so one can  say  that  the  region  of  the  lowest
temperatures is fluctuationless.

Turning back to the inequality (\ref{e68}) and taking into account
the estimation (\ref{e71}) one can conclude that the region of importance of
thermal fluctuations at least for temperatures above $T\sim \Gamma$ 
presented by (\ref{e70}) is given  by  inequality  (\ref{e68})  that  is
broader  than  the  period   of   critical   temperature   oscillations
(\ref{e31}):
$$
\frac{\delta H}{H_{c2}}\approx2\pi^2\left(\frac{T_c}{\varepsilon_F}\right)^2
$$
Thus in this temperature region the oscillations of critical temperature
have poor chances for observations. On the other  hand at  temperatures
\begin{equation}
T<\Gamma<\frac{\omega_{c2}}{4\pi}
\label{e73}
\end{equation}
we have  region  free  of  fluctuations
where one  can  hope  to  find  the  quantum  oscillations  of  critical
temperature.
\section {Conclusion}
The quantitative   theory   presented  in  the  paper  demonstrates  the
suppression of  the  de  Haas  van  Alphen  effect  in  the
superconducting mixed state  according the equations (\ref{e59}),
(\ref{e60}).
The suppression is developed in the region below the upper critical field 
embracing  up  to  hundred
oscillations of magnetization  (\ref{e63}) and limited by the  
inequality  (\ref{e57}). 
The quantum oscillatons of the critical temperature are accessible
for observation at very low temperatures (\ref{e73}).
\section*{Acknowledgments}
It was happened that the work under this paper falled on the time
of finishing of my responsibility for the Landau Institute and 
the commencement of  the new duties at Comissariat a l'Energie Atomic.
I am indebted to Jacques Flouquet for his permanent kind care
as well to all my friends for  the emotional  support  during
this time.



\begin{thebibliography}{99}
\bibitem[$^{*)}$]{*)}E-mail:mineev@drfmc.ceng.cea.fr

\bibitem{1} T.J.B.M.Janssen,  C.Haworth,  S.M.Hayden, P.Meeson 
and M.Springford,  Phys. Rev. B {\bf 57},11698 (1998).
 
\bibitem{2} H.Ohkuni,  T.Ishida, Y.Inada, Y.Haga, E.Yamamoto, Y.Onuki, 
S.Takahashi, Journ. Phys. Soc. Jap.{\bf 66}, 945 (1997).

\bibitem{3} T.Terashima,  C.Haworth, H.Takeya, S.Uji and H.Aoki, 
Phys. Rev. {\bf 56}, 5120 (1997).

\bibitem{4} C.Bergemann,  S.R.Julian, G.J.McMullan, B.K.Howard, 
G.G.Lonzarich, P.Lejay, J.P.Brison, J.Flouquet, 
Physica B {\bf 230-232},438 (1997).

\bibitem{5} K.Maki, Phys, Rev. B {\bf 44}, 2861 (1991).

\bibitem{6} M.J.Stephen, Phys. Rev. B {\bf 45}, 5481 (1992).

\bibitem{7} K.Miyake, Physica B {\bf 186},115 (1993).

\bibitem{8}S.Dukan and Z.Tesanovic, Phys. Rev. Lett. {\bf 74},2311 (1995).

\bibitem{9}P.Miller and B.L.Gyorffy, 
J.Phys.: Condens. Matter {\bf 7}, 5579 (1995).

\bibitem{10}M.R.Norman and A.H.MacDonald, Phys.Rev.B {\bf 54}, 4239 (1996).

\bibitem{11}G.M.Bruun ,  
V.N.Nicopoulos and N.F.Johnson, Phys.Rev.B {\bf 56}, 809 (1997).

\bibitem{12}V.N.Zhuravlev, T.Maniv, I.D.Vagner, P.Wyder, 
Phys. Rev. B {\bf 56},14693 (1997).

\bibitem{13}M.G.Vavilov and V.P.Mineev, Zh. \'Eksp. Teor. Fiz. {\bf 112},
1873 (1997) [Sov.Phys. JETP {\bf 85},1024 (1997)].

\bibitem{14}V.P.Mineev, Physica B, {\bf 259}-{\bf 261}, 1072 (1999).

\bibitem{15}L.P.Gor'kov and J.R.Schrieffer, Phys.Rev.Let., 
{\bf 80}, 3360 (1998);
Preprint cond-mat/9810065, 1998.

\bibitem{16}L.W.Gruenberg and L.Gunther, Phys. Rev. {\bf 176}, 606 (1968).

\bibitem{17}L.P.Gor'kov, Zh. \'Eksp.  Teor. Fiz. {\bf 37}, 833 (1959).

\bibitem{18}I.M.Lifshits, I.A.Kosevich,
Zh.\'Eksp.  Teor. Fiz. {\bf 29}, 730 (1955) 

\bibitem{19}V.P.Mineev, Proc. of Int. Conference on "Physical Phenomena
at High Magnetic Fields", Oct.1998, Tallahassee, Florida, World Scientific
(1999).


\bibitem{20}E.Brezin, A.Fujita, S.Hikami, Phys. Rev. Lett. {\bf 65}, 
1949 (1990).

\bibitem{21}C.Caroli, K.Maki, Phys. Rev. {\bf 159}, 316 (1967).


\end{thebibliography}
\end{document}